# Concept of a novel fast neutron imaging detector based on THGEM for fan-beam tomography applications

**M. Cortesi,**[a*] **R. Zboray,**[a] **R. Adams,**[a,b] **V. Dangendorf**[c] **and H.-M. Prasser**[a,b]

[a] *Paul Scherrer Institut (PSI),*
*Villigen PSI 5232, Switzerland*

[b] *Swiss Federal Institute of Technology (ETH)*
*Zürich 8092, Switzerland*

[c] *Physikalisch-Technische Bundesanstalt (PTB),*
*Braunschweig 38116, Germany*
*E-mail*: marco.cortesi@psi.ch

ABSTRACT: The conceptual design and operational principle of a novel high-efficiency, fast neutron imaging detector based on THGEM, intended for future fan-beam transmission tomography applications, is described. We report on a feasibility study based on theoretical modeling and computer simulations of a possible detector configuration prototype. In particular we discuss results regarding the optimization of detector geometry, estimation of its general performance, and expected imaging quality: it has been estimated that detection efficiency of around 5-8% can be achieved for 2.5MeV neutrons; spatial resolution is around one millimeter with no substantial degradation due to scattering effects. The foreseen applications of the imaging system are neutron tomography in non-destructive testing for the nuclear energy industry, including examination of spent nuclear fuel bundles, detection of explosives or drugs, as well as investigation of thermal hydraulics phenomena (e.g., two-phase flow, heat transfer, phase change, coolant dynamics, and liquid metal flow).

KEYWORDS: Gas Electron Multiplier; THGEM; Neutron detector; Neutron Radiography, Neutron Tomography;

---

[*] Corresponding author.

# Contents

## 1. Introduction

Fast neutron radiography and tomography are powerful tools for non-destructive analysis (NDA)of samples and for imaging of technological processes, applicable to a broad field of industrial and basic research applications [1]. The special features of neutron interaction with matter, together with their unique material penetrating properties, allow for inspection of bulk specimen and production of images of light elements (such as hydrogen) beneath a matrix of metallic elements. In addition, the possibility of investigating fast time-dependent phenomena has particularly important technological and industrial applications, and is a wide and relatively unexplored area of research.

Particularly relevant for nuclear power technology R&D is the investigation of thermal-hydraulic phenomena such as two-phase flows in the context of boiling water reactor fuel bundle studies. In the past, X- and gamma-ray radiography and tomography have been the first choice among the non-invasive imaging techniques, both for general two-phase flow research and for nuclear fuel bundle investigation [2]. More recently, imaging techniques such as thermal-neutron radiography [3] and three-dimensional neutron tomography [4] were also used for investigation of liquid-metal two phase flows in the context of generation IV reactors [4, 5] and corium-water thermal interactions (e.g., steam explosions) [6, 7].

All of the aforementioned radiographic imaging techniques are performed in a steady beamline. In the case of tomographic analysis, the reconstructed image is obtained by rotating the object to obtain different projection angles, without any possibility of time-resolved analysis. High frame rate imaging capability has been achieved using a scanned electron beam X-ray tomography at the Rossendorf Ultrafast Electron Beam X-ray Tomography (ROFEX) facility, capable of producing up to 10000 images/second of steady specimen at a spatial resolution of about 1 mm [8]. In that device the beam from an electron gun, accelerated up to 150kV, is focused on, and very quickly swept across, a tungsten target by using an electromagnetic deflection system, thereby producing a moving X-ray emitting spot. A static ring of solid-state X-ray detectors is placed around the object. Using neutrons in a similar fashion could enable the imaging of dynamic processes in robust, attenuating specimen that are otherwise opaque for X-ray or gamma photons. For such a setup only fast neutrons could potentially come into consideration [9].

Our research activities are focused on the development of a fan-beam neutron tomography system (source-detector combination), capable of producing 2D cross-sectional images of non-rotating objects, suitable for studying two-phase flows. The present research project aims at development of a small-scale tomographic system prototype in order to prove the concept, provide expected imaging performance, and evaluate the feasibility of adding high-frame-rate capability.

A compact, fast neutron generator with a small (mm size) emitting spot is currently under development by our group. It is an accelerator-based neutron source, using deuterium-deuterium fusion reactions to produce fast neutrons of 2.45MeV. Deuterium ions are extracted from an RF-driven plasma ion source and accelerated up to 150 keV, hitting a Ti target fixed on an actively cooled copper rod. Detailed information about development of this fast neutron source prototype will be shortly reported elsewhere [9].

In this work we present and discuss the conceptual design and operational principle of a novel fast neutron imaging detector suitable as a fan-beam tomography readout, based on THick Gaseous Electron Multiplier (THGEM) [10, 11]. In addition, we report on the results of a systematic computer simulations study which provides detailed information about expected performance and imaging capability of the proposed imaging detector system. Prospects and potential applications are also discussed.

## 2. Conceptual design of the fan-beam tomography system

The conceptual design of the fan-beam fast neutron tomographic system is depicted in figure 1a. The system is comprised of many fast neutron point-like sources, sequentially pulsed, and placed around a not-rotating object. A ring-shaped imaging detector provides the projection images of the investigated object.

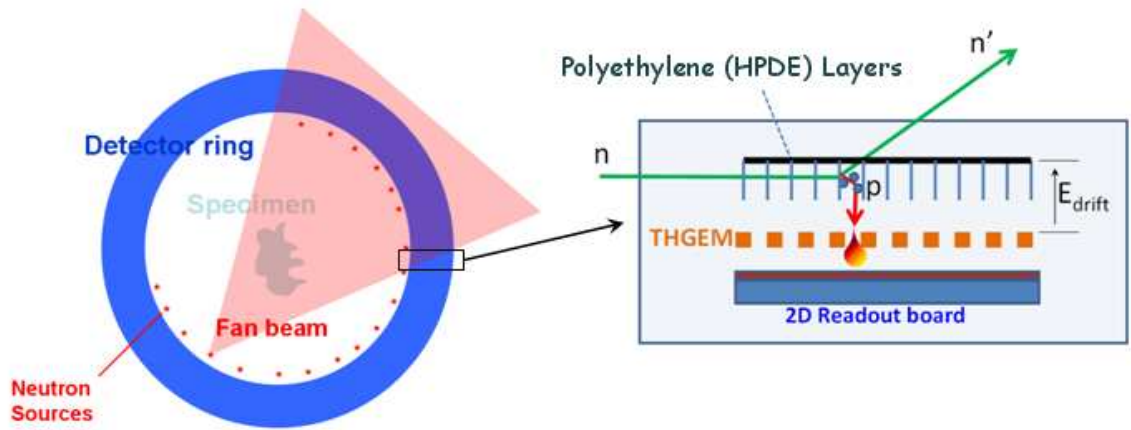

**Figure 1.** Part a: conceptual design of the high-frame-rate fast neutron tomographic system. Part b: schematic drawing of a vertical section of the imaging detector based on a THGEM-based readout.

The fast neutron detector, proposed here, consists of a stack of conductive polyethylene foils immersed in a gas medium and coupled to a THGEM-based readout. In this approach the foils are arranged perpendicularly to the direction of the impinging neutron beam, as shown in figure 1b. Collision processes (mainly elastic n-p scattering) may occur between impinging neutrons and hydrogen nuclei in one of the many converter foils along the neutron direction; if the resulting recoil protons have enough energy, they can escape the converter foil and ionize the gas medium in between two successive converters. Upon application of a suitable electric field (around 1 kV/cm), the ionization electrons are drifted, parallel to the converter-foil surfaces, towards the two-dimensional arc-shaped THGEM-based readout, for gas avalanche multiplication and localization. The one-dimensional projection image of the investigated object corresponds to the one-dimensional distribution of neutron attenuation inside the object, integrated over the projection chords. The cross-sectional tomographic image of the object can then be reconstructed based on the projection images, recorded at different angles.

## 3. THGEM-based Imaging Detector

The building block of the imaging detector is the THGEM, a novel hole-type gaseous electron multiplier structure. It is fabricated using standard PCB techniques, consisting of perforated submillimeter holes in a double metal-clad FR4 plate; the production process is terminated by a chemical etching of the rim around each hole, which is essential for reducing discharges which could be triggered by mechanical defects. Each hole functions as an independent proportional counter: upon application of a voltage difference across the THGEM, a strong dipole electric field is established within the holes. This strong field (few tens kV/cm) is responsible for an efficient focusing of ionization electrons into the holes, and their multiplication by gas avalanche processes. It is also possible to cascade several elements and to obtain higher detector gain at lower operating voltage per electrode. For a concise review of the operation and proprieties of THGEM the reader is referred to [12] and references therein.

Although THGEMs can operate in a large variety of gases, providing a high electron multiplication factor (up to $10^4$-$10^5$ in a single element and $10^6$-$10^7$ with two THGEM in cascade), the operation of Ne and Ne-based mixtures is of particular interest: it provides high electrons multiplication at very low operational voltages compared to other standard gas mixtures (for example Ar-based mixture) [13]; low operational voltage has the advantage of providing more stable operational conditions and lower probability of damaging the electrode following electric discharges.

Most importantly, Ne-based mixtures provide a large dynamic range [13]; indeed, as a consequence of the high electron diffusion coefficient, which is characteristic of these gas mixtures, the electron-avalanche is extended over a large volume, and thus it is possible to build up a considerable amount of charge before reaching the space charge density limit. Large dynamic range is particularly crucial for applications with a highly-ionizing radiation background and for applications with a wide spectrum of deposited energy, which is always the case in fast-neutron interaction with matter therein.

The localization properties of THGEM-based imaging detectors, operating in various gas mixtures (Ar- and Ne-based mixtures), were studied with soft (< 10 keV) X-rays. A sub-millimetre spatial resolution (FWHM) was obtained with good linearity and good homogeneity across the whole detector active area. For more info about these studies the reader is referred to the following work [14] and references therein.

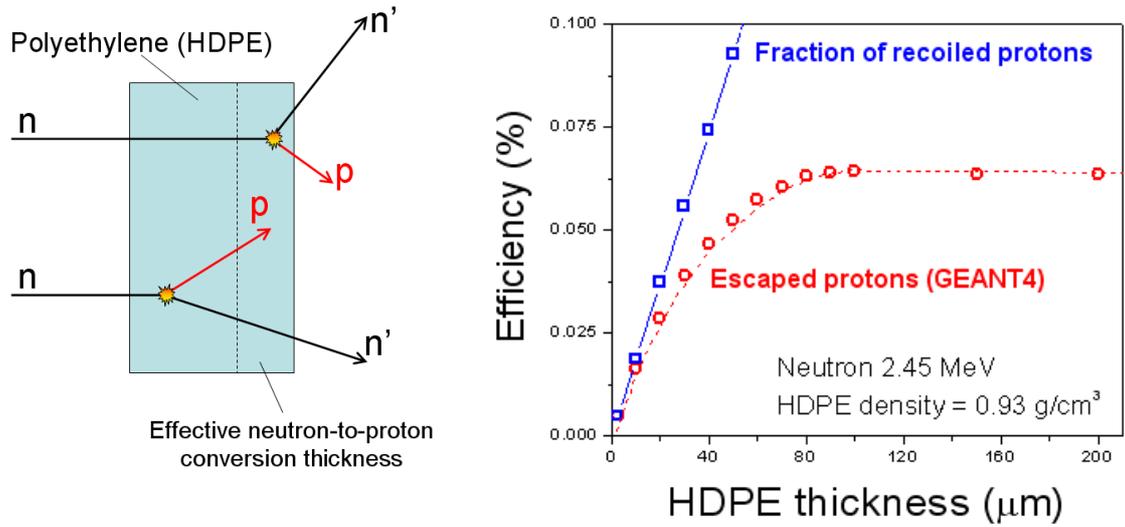

**Figure 2.** Red symbols: detection efficiency as a function of the converter thickness. Blue symbols: neutron-to-proton conversion efficiency as a function of the converter thickness. The converter foils are made of polyethylene (mass density of 0.93 g/cm$^3$).

## 4. Neutron-to-proton conversion using a stack of polyethylene foils

### 4.1 Thickness of the polyethylene foils

The detection efficiency of this device depends both on the elastic scattering cross section and the probability of the neutron-induced protons to escape the converter foils (figure 2a). The latter is related to the range of the protons in the material from which the converter is made (in our case polyethylene). Figure 2b illustrates the fraction of neutrons converted into an escaping recoil proton, as a function of the converter thickness (red graph); this computation was performed by a series of GEANT [15] computer simulations assuming an impinging monoenergetic neutron beam of 2.45MeV. As shown in figure 2b, the detection efficiency initially increases linearly with the converter thickness, and then saturates (at a value of around 0.07%) when the converter thickness approach the range of the most energetic recoil protons. For forward scattered recoil protons of 2.45MeV in polyethylene (with mass density of r = 0.93 g/cm3), the range is around 100 mm.

For comparison, the conversion efficiency (namely the total fraction of neutrons interacting in the converter foil) is also shown in figure 2b. From the non-relativistic kinematic analysis of the elastic scattering process, it results that the kinetic energy of the recoil target nucleus (ER), induced by monoenergetic neutrons (En), is given by the following relation::

$$E_R = \frac{4A}{(1+A)^2} \cos^2(\theta) E_n \qquad \text{(Eq. 1)}$$

where A is the mass of the target (A=1 for proton) and θ is the scattering angle in the laboratory coordinate system. For relative thin converter foil (the order of the saturation thickness), the recoil proton energy distribution is approximately rectangular, extending from zero to the full incident neutron energy (figure 3).

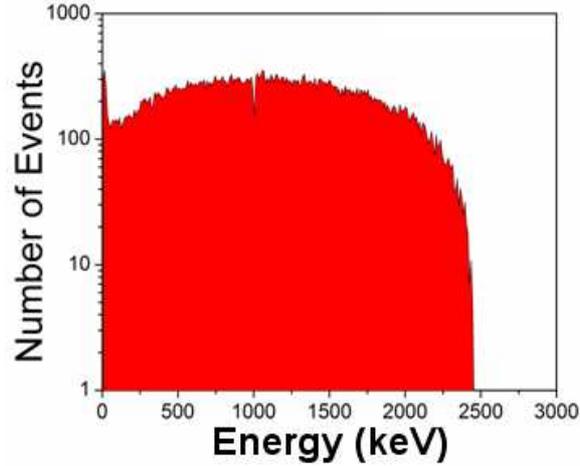

**Figure 3.** Energy distribution of neutron-induced recoil protons escaping the converter foil.

## 4.2 Energy deposited in the gas medium between two converter foils

The recoil protons escaping the converter create ionization electrons in the gas medium (Ne-based mixture) between the emitting converter and the successive one (the red tracks in figure 4a corresponds to delta electrons produced along the proton path - blue track). The energy deposited by the proton in to the gas is only a small fraction of its total kinetic energy (the distribution of which is shown in figure 2 and depends on the gap length between the two converters; the larger the distance between the two consecutive converters, the larger the amount of energy deposited by the recoil proton. According to GEANT4 simulations, in the range between 300 mm and 700 mm the average value of the deposited energy ranges from 4.4 to 9 keV (figure 4b).

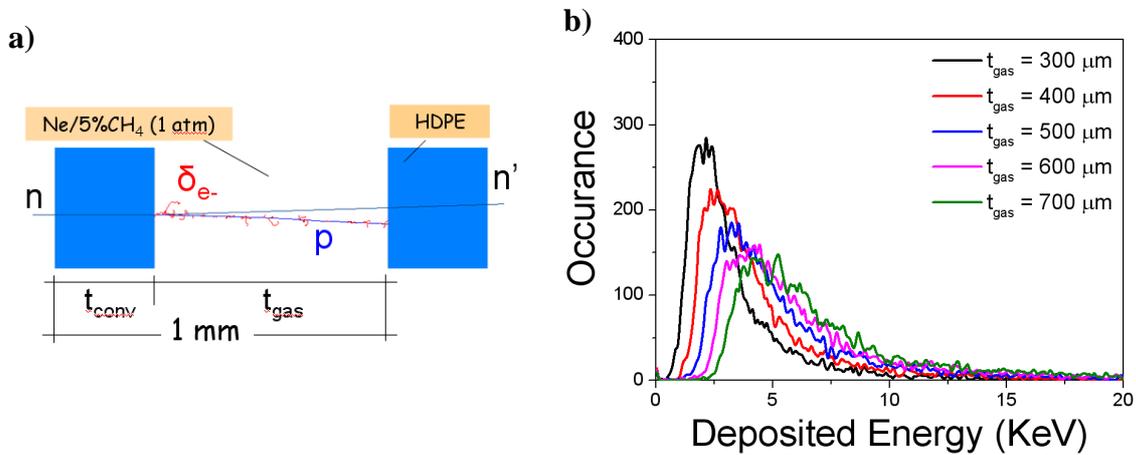

**Figure 4.** Part a) Snapshot of a GEANT4 simulation run: it shows a neutron (green track) scattered in a converter foil (made of polyethylene – HDPE); as result of this interaction, a recoil proton (blue track) is emitted from the converter and ionizes (red tracks are visualization of delta electrons) the gas in the gap between the two foils. The proton is then stopped in the successive converter foil. Part b) Resulting distribution of the energy deposited in the gas gap by recoil protons ($t_{gas}$ is the width of the gas gap).

In terms of detector performance, the most important feature is the spread of the deposited energy in the gas, since this quantity will pose a demand on the dynamic range of the detector. As shown in the distributions of figure 4b, the widths (FWHM) of the deposited energy distributions are from 5 to 10 KeV, depending on the width of the gas gap: the distributions are somewhat broadened for wider gas gaps. As discussed in section 3, large electron multiplication factors with a large dynamic range could be achieved using THGEM operating in Ne-based mixtures, due to broader electron-avalanche spread. This should guarantee the possibility of processing signals with a broad pulse-height distribution and to achieve the highest detection efficiency. It should be noted that only n-p interactions are considered here. Events caused by recoiling carbon ions, which can cause much larger energy deposition in the gas, are neglected because their occurrence is rare. Nevertheless, such events can ignite sparks in a gas detector and their handling (e.g. by quenching) must be studied in the future.

**4.3 Multi-layer converter**

As discussed in the previous paragraphs, the greatest conversion efficiency is achieved when the thickness of the converter foil is at least equal of the range of the most energetic recoil protons (100 mm); in addition, the gas gap between the converter should be as small as possible so as to reduce the spread of the deposited energy by the recoil proton (ideally 300 mm). Due to limitations of the present production technique and mechanical constraints (fragility), it is not possible to produce structures comprising of many extended foils (few hundreds) with such a small thickness (few hundred mm) and tight configuration (100 foils per cm). Presently, novel manufacturing technologies like three-dimensional printing [16] allow three-dimensional objects to be created by raising successive layers of it out of various materials (acrylonitrile butadiene styrene - ABS, polyethylene - HDPE, etc.) in quite fine detail. In the present configuration, as a first stage, we managed to produce a converter with both a foil and gap width of 600 mm.

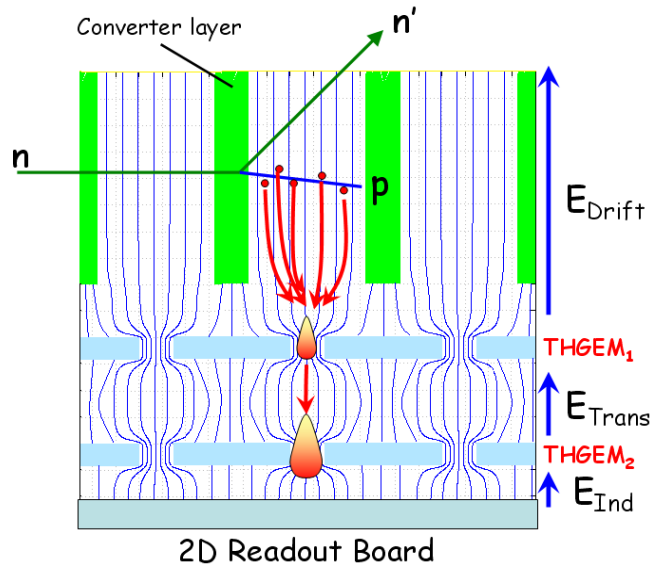

**Figure 5.** Concept of the fast neutron detector based on THGEM readout.

Upon application of a suitable electric field (around 1 kV/cm), the ionization electrons, induced by the recoil protons in the gas medium, are drifted parallel to the foil surfaces towards the THGEM-based readout for detection and localization (figure 5). However, the ratio between the drift region of the electron in the converter stack (i.e the height of the converter h = 6 mm) is several times the width of the gas gap between foils ($t_{gas}$ = 0.6 mm). For such a configuration it is very likely that even small charging up of the polyethylene, from which the converters are made of, may cause a distortion of the electric field in the drift region. This affects the efficiency in collecting and focusing the ionization electrons into the THGEM holes and thus may significantly decrease the overall detection efficiency.

To avoid any loss of efficiency due to charging up of the converter, we plan to use converter foils made of low-resistivity (anti-static/dissipative) polyethylene and to apply a voltage between the upper and lower face of the converters structure; in that way one can maintain a constant (and undistorted) electric field in the drift region, removing the charge from the surface of the converter.

### 4.3.1 Detection efficiency

Figure 6 illustrates the calculation of the detection efficiency as a function of the total number of the converter foils in the stack; these results were obtained by GEANT4 simulations that assumed converter foils with a thickness of 300 μm and a gas gap of 300 μm width. At the beginning, the detection efficiency linearly increases with the number of converter foils, and it saturates at a value of around 8% when the stack of converter foils reaches around 300 units. At this point the impinging neutron beam is extinguished by scattering processes: all the neutrons interacting ahead of the effective converter thickness (namely before the last 100 μm of the converter), cannot be detected - their induced recoil protons do not escape the converter.

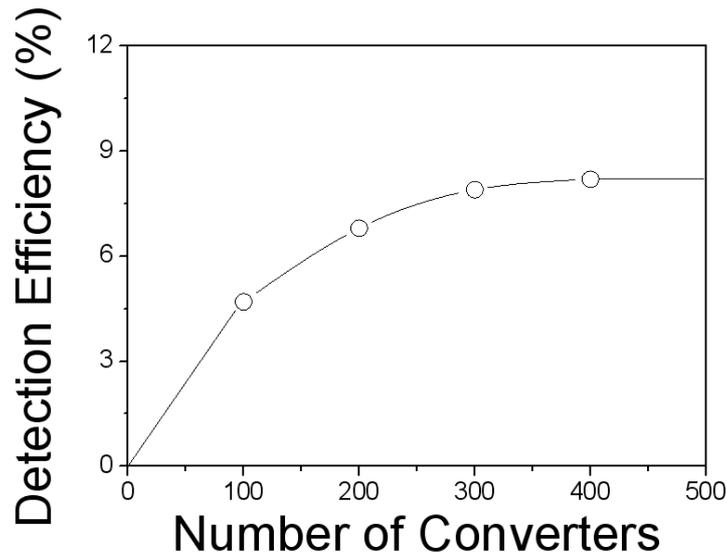

**Figure 5.** Detection efficiency as function of number of converter foils (converter thickness = 300 μm, gas gap = 300 μm).

### 4.3.2 Spatial resolution

The spatial resolution of the imaging detector ($SR_{tot}$) depends on various parameters:

$$SR_{tot} = SR_{Conv.} * SR_{Diff.} * SR_{Readout} \qquad (Eq.\ 1)$$

where the symbol * stands for convolution. $SR_{Conv.}$ is the contribution corresponding to angular spread of the primary charge induced by recoil proton in the gas medium. This contribution is negligible for recoil protons that are scattered forward with the highest recoil energy; they have high probability to escape the converter foil and to deposit their energy in a well localized area in the gas medium between two successive converters. On the contrary, protons that are scattered at large angles and may deposit energy in a large area in the gas medium, spoiling the spatial resolution, have low kinetic energy and usually do not escape from the converter surface. The factor $SR_{Diff}$ takes into account the spread of the ionization electrons due to diffusion when they drift towards the THGEM electrodes. The drift region and the spread of the electron-avalanche are rather small and thus this contribution is generally negligible. Eventually, $SR_{Readout}$ is determined by the contributions to the spatial resolution of both, the induced charge to the pick-up readout electrode through a resistive anode and the degradation of the spatial resolution by signal processing in the front-end electronics. The quantity ($SR_{Diff.}*SR_{Readout}$) is an intrinsic characteristic of the detector readout. Reference [14] reported sub-millimeter position resolution (around 700 μm) obtained with a THGEM-based detector prototype tested with few-keV X-rays.

The distribution of the charge induced by the recoil protons in the gas medium ($SR_{Conv}$), along the multi-layer converter structure, as a function of the number of foils in the converter structure (100, 200, 300 foils), is depicted in figure 6. In the simulations it was assumed that a mono-directional, mono-energetic (2.45 MeV) and infinitesimal small neutron beam is impinging on the center of a multi-layer structure, comprised of foils of dimension: 100x100x5 mm.

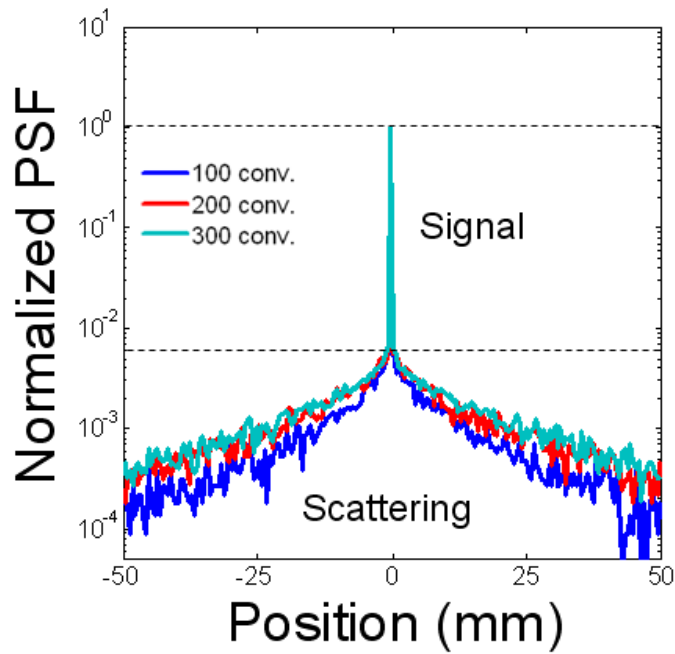

**Figure 6.** Spatial distribution of the proton-induced charge created in the gas medium as function in various number of converter (100, 200 and 300 μm).

The resulting spatial distribution (named as point spread function – PSF, and normalized by the maximum value of the distribution) of the induced charge in figure 6, shows a sharp peak (around 500 μm FWHM) corresponding to the interaction of impinging neutron from the uncollided beam. The peak is on the top of a tiny flat continuum originating from multiple scattering of neutrons inside the many foils of the converter.

## 5. Discussion and Conclusion

The conceptual design and operational principle of a novel fast neutron imaging detector, intended for future fan-beam transmission tomography applications, is presented and described. The detector is comprised of a multi-layer converter for neutron-to-proton conversion, coupled to THGEM for proton-induced charge localization. The optimization and characterization of multi-layer converters made of polyethylene were investigated by means of systematic Monte Carlo simulations using GEANT4. It was calculated that an optimal and manufacturable configuration consists of 300-400 units, each 300 μm thick and separated by 300 μm; in this case a detection efficiency on the order of 8% and a spatial resolution of the tomographic reconstructed image of around 1 mm are foreseen; the degradation of the image contrast due to neutron multi-scattering inside the stack of converters may be corrected using standard image processing.

As briefly discussed in section 4.3, one possible concern is the electrostatic stability of the field between the converter foils. Charging-up of the polyethylene foils may deform the geometry of the electric field in the drift region, with the consequence of losing the capability of efficiently drifting the ionization electrons towards the THGEM-based readout, causing lose of detection efficiency. Currently, we have testing the performance of a multi-layer converter prototype, made of static dissipative polyethylene, manufacture by 3D printing. The idea is to apply a different voltage between the upper and lower surface of the converters stack, such that one can maintain a constant (and undistorted) electric field in the drift region, preventing a build-up of static charge on the surface of the foils. Particular attention has been devoted to tailor the volume resistivity of the polyethylene. On the one hand, too low resistivity produces too high electrical energy dissipation that may increase the temperature of the foils, compromising their mechanical stability and changing the operating conditions of the filling gas in the drift region. On the other hand, if too isolating, they may charge up. The prototype will be characterized through a series of systematic laboratory tests and the results will be reported later.

The proposed instrument may find many important technological and industrial applications: dynamic visualization of the combustion engine fluid dynamics for studying the optimization of engine performances; non-destructive monitoring of capillary processes, such as in water transport in porous building materials; investigation of heat exchange of fluidized-bed heat exchangers in steel industry; investigations of turbulent oil-gas flow through a pipe in petrochemical industry. Particularly important is the study of phenomena relevant for development of nuclear plant technologies, including: investigation of (gas-liquid or gas-solid) two-phase flow; study of steam explosion processes, such as in severe accident of a nuclear reactor due to the direct contact of molten core and coolant and finally nuclear fuel inspection.

## Acknowledgments

R. Adams is supported by the Swiss National Science Foundation (SNSF) grant number: 200021-12987C/1.

**References**

[1]  D. Vartsky, *Prospects of Fast-Neutron Resonance Radiography and its Requirements for Instrumentation*, proceeding of the International Workshop on Fast Neutron Detectors and Applications (FNDA2006), Cape Town, South Africa, 3 - 6 Apr 2006, *PoS(FNDA2006)*, pg. 84.  http://pos.sissa.it/archive/conferences/025/084/FNDA2006_084.pdf

[2]  B. Neykov, O. N. E. Agency, N. E. A. N. S. Committee, and N. E. A. C. on S. of N. Installations, *NUPEC BWR full-size fine-mesh bundle test (BFBT) benchmark: Specifications*. Nuclear Energy Agency, Organisation for Economic Cooperation and Development, 2006.

[3]  M. Kureta, *Experimental Study of Three-Dimensional Void Fraction Distribution in Heated Tight-Lattice Rod Bundles Using Three-Dimensional Neutron Tomography*, *Journal of Power and Energy Systems*, vol. 1, no. 3, pp. 225-238, 2007.

[5]  Y. Saito et al., *Application of high frame-rate neutron radiography to liquid-metal two-phase flow research*, *NIM A* **542** (2005), pp. 168-174.

[6]  Y. Saito, K. Mishima, Y. Tobita, T. Suzuki, and M. Matsubayashi, *Measurements of liquid–metal two-phase flow by using neutron radiography and electrical conductivity probe*, *Experimental Thermal and Fluid Science* **29** (2005), pp. 323-330.

[7]  Y. Saito, K. Mishima, T. Hibiki, A. Yamamoto, J. Sugimoto, and K. Moriyama, *Application of high-frame-rate neutron radiography to steam explosion research*, *NIM A* **424** (1999), pp. 142-147.

[8]  Y. Sibamoto, Y. Kukita, and H. Nakamura, "Visualization and Measurement of Subcooled Water Jet Injection into High-Temperature Melt by Using High-Frame-Rate Neutron Radiography," *Nuclear Technology* vol. **139** (2002), pp. 205-220.

[8]  F. Fischer and U. Hampel, "Ultra fast electron beam X-ray computed tomography for two-phase flow measurement," *Nuclear Engineering and Design*, vol. 240, no. 9, pp. 2254-2259, Sep. 2010.

[9]  R. Zboray, *Development of a compact D-D neutron generator with a small emitting spot for fast neutron imaging*, oral contribution to the International Workshop on Fast Neutron Detectors and Applications (FNDA2011), Ein Gedi, Israel, Nov-2011.

[10]  C. Shalem, R. Chechik, A. Breskin, and K. Michaeli, *Advances in Thick GEM-like gaseous electron multipliers. Part I: atmospheric pressure operation*, *NIM A* **558** (2006), pp. 475.

[11]  R. Chechik, A. Breskin, C. Shalem, and D. Mörmann, *Thick GEM-like hole multipliers: properties and possible applications*, *NIM A* vol. **535** (2004), pp. 303-308.

[12]  A. Breskin et al., *A concise review on THGEM detectors*, *NIM A* vol. **598** (2009), pp. 107.

[13]  M. Cortesi et al., *THGEM operation in Ne and Ne/CH4*, *JINST* vol. **4** (2009), pp. P08001-P08001.

[14]  M. Cortesi et al., *Investigations of a THGEM-based imaging detector*, *JINST*, vol. **2** (2007), pp. P09002-P09002.

[15]  S. Agostinelli et al., *Geant4—a simulation toolkit*, *NIM A* **506** (2003), pp. 250-303.

[14]  L. M. Sherman, (2004), http://www.ptonline.com.